\documentclass[journal,11pt,draftclsnofoot,onecolumn]{IEEEtran}

\usepackage[noadjust]{cite}

\usepackage[T1]{fontenc}
\usepackage{ae}
\usepackage{aecompl}

\ifCLASSINFOpdf
\else
\fi
\usepackage[cmex10]{amsmath}
\usepackage{amssymb}
\newtheorem{definition}{Definition}[section]
\newtheorem{theorem}{Theorem}[section]
\newtheorem{lemma}{Lemma}[section]
\newtheorem{remark}{Remark}[section]

\newtheorem{corollary}{Corollary}[section]

\begin{document}
%
\title{Overflow Probability of Variable-length Codes with Codeword Cost}
%
%
%

\author{Ryo Nomura, \IEEEmembership{Member, IEEE}
\thanks{R. Nomura is with the School of Network and Information, Senshu University, Kanagawa, Japan,
 e-mail: nomu@isc.senshu-u.ac.jp. 
The material in this paper was presented in part \cite{Nomura_ISIT2012a} at the 2012 IEEE International Symposium on Information Theory, Boston, USA, July 2012.}
\thanks{The first author with this work was supported in part by JSPS Grant-in-Aid for Young Scientists (B) No. 23760346.}
}

\maketitle

\begin{abstract}
Lossless variable-length source coding with codeword cost is considered for {\it general} sources.
The problem setting, where we impose on unequal costs on code symbols, is called the variable-length coding with codeword cost.
In this problem, the infimum of average codeword cost have been determined for {\it general} sources.
On the other hand, overflow probability, which is defined as the probability of codeword cost being above a threshold, have not been considered yet. 
In this paper, we determine the infimum of achievable threshold in the first-order sense and the second-order sense for general sources and compute it for some special sources such as i.i.d. sources and mixed sources.
A relationship between the overflow probability of variable-length coding and the error probability of fixed-length coding is also revealed.
Our analysis is based on the {\it information-spectrum} methods.
\end{abstract}


%
\IEEEpeerreviewmaketitle

%
%
%
%


\section{Introduction}
\IEEEPARstart{L}{ossless} variable-length coding problem is quite important not only from the theoretical viewpoint but also from the viewpoint of its practical applications.
To evaluate the performance of variable-length codes, several criteria have been proposed.
The most fundamental criterion is the average codeword length, which is proposed by Shannon \cite{Shannon}.
And then, many variable-length codes have been proposed and its performance has been evaluated  by using the average codeword length \cite{Cover}.
The overflow probability of codeword length is one of other criteria, which denotes the probability of codeword length per symbol being above a threshold $R >0$.
Merhav  \cite{Merhav91:overflow} has first determined the optimal exponent of the overflow probability given $R$ for unifilar sources.
Uchida and Han \cite{Uchida01:overflow} have shown the infimum of achievable threshold $R$ under constraints that the overflow probability vanishes with given exponent $r$.
Their analyses are based on the {\it information-spectrum} methods and the results are valid for {\it general} sources. 
Nomura and Matsushima \cite{Nomura2011} have computed the infimum of $\varepsilon$-achievable overflow threshold for {\it general} sources. Here, $\varepsilon$-achievable overflow threshold means that there exists a variable-length code, whose overflow probability is smaller than or equal to $\varepsilon$.
All the results mentioned here are in the meaning of the codeword length.

On the other hand, as is well known, if we impose {\it unequal} costs on code symbols, 
it makes no sense to use the codeword length as a measure. 
In this setting, we have to consider the codeword {\it cost}, instead of codeword length.
The average codeword {\it cost}, which is a generalization of the average codeword length, have been first analyzed also by Shannon \cite{Shannon}.
Moreover, several researchers have studied on the average codeword cost \cite{Csiszar69:cost,CK,Krause}.
Karp \cite{Karp} has given the variable-length code, which minimizes the average codeword cost. 
Golin and Rote \cite{Golin98:cost}, and Golin and Li \cite{Golin2008} have proposed the efficient algorithm constructing the optimal variable-length code for i.i.d. sources and Iwata, Morii and Uyematsu \cite{Iwata97:cost} have proposed the asymptotic optimal universal variable-length code for stationary ergodic sources, with respect to the average codeword cost.
The infimum of the average codeword cost has been determined by Krause \cite{Krause} for i.i.d. sources and extended to {\it general} sources by Han and Uchida \cite{Han00:cost}.
Among others, Uchida and Han \cite{Uchida01:overflow} have proposed the overflow probability of codeword {\it cost}. They have considered the overflow probability as the probability of codeword {\it cost} per symbol being above a threshold. Then, they have shown the infimum of achievable threshold, where achievable threshold means that there exists a variable-length code whose overflow probability of codeword cost decreases with given exponent $r$.

In this paper, we also deal with the overflow probability of codeword {\it cost}.
In particular, we consider the $\varepsilon$-achievable threshold, which means that there exists a variable-length code, whose overflow probability is smaller than or equal to $\varepsilon$.
We first reveal the relationship between the overflow probability of variable-length coding and the error probability of fixed-length coding.
Second, we determine the infimum of {\it first-order} and {\it second-order} achievable threshold for {\it general} sources.
The finer evaluation of the achievable rate, called the second-order achievable rate, has been investigated in several contexts. In the variable-length source coding problem, Kontoyiannis \cite{Kon} has established the second-order source coding theorem on the codeword length for i.i.d. sources and Markov sources. 
In the channel coding problem, Strassen \cite{Strassen} (see, Csisz\"{a}r and K\"{o}rner \cite{CK}), Hayashi \cite{Hayashi2}, and Polyanskiy, Poor and Verd\'{u} \cite{Poly2010} have determined the second-order capacity rate. Hayashi \cite{Hayashi} has also shown the second-order achievability theorems for the fixed-length source coding problem for general sources and compute the {\it optimal} second-order achievable rates for i.i.d. sources by using the asymptotic normality.
Nomura and Han \cite{NH2011} have also computed the optimal second-order achievable rates in fixed-length source coding for {\it mixed} sources by using the two-peak asymptotic normality.

Analogously to these settings, we define the {\it second-order} achievable threshold on the overflow probability and derive the infimum of the {\it second-order} achievable threshold. 
Notice here that Nomura and Matsushima \cite{Nomura2011} have already considered the first-order and the second-order achievability with respect to the overflow probability of {\it codeword length}. 
One of contributions of this paper is a generalization of their results into the case of {\it codeword cost}.
Our analysis is based on the {\it information-spectrum} methods and hence our results are valid for {\it general} sources.
Furthermore, we apply our results to i.i.d. sources and mixed sources as special cases and compute the infimum of the {\it second-order} achievable threshold for these special but important sources.

Related works include works by Kontoyiannis and Verd\'{u} \cite{KV2013}, and Kosut and Sankar \cite{KS2013}. They have also considered the similar quantity with the overflow probability.
Kontoyiannis and Verd\'{u} \cite{KV2013} have derived the fundamental limit of this quantity without the prefix conditions.
Kosut and Sankar \cite{KS2013} have also derived the upper-bound of the overflow probability in universal setting.
It should be emphasized that they have considered the overflow probability of codeword {\it length} for some special sources and derived bounds up to the third-order. On the other hand, in this paper we have considered the overflow probability of codeword {\it cost} for {\it general} sources and addressed the fundamental limit of the achievable threshold up to the second-order.

This paper is organized as follows.
In Section II, we state the problem settings and define the achievability treated in this paper.
In Section III, we reveal the relationship between the overflow probability of variable-length coding and the error probability of fixed-length coding. 
In Section IV, we prove two lemmas which play the key role in the subsequent analysis.
In Section V, we determine the infimum of {\it first-order} achievable threshold.
In Section VI, we derive the infimum of {\it second-order} achievable threshold and compute it for some special sources.
In Section VII, we conclude our results. 
\section{Overflow probability of Variable-length Coding with cost}
\subsection{Variable-length codes with codeword cost for general source}
The {\it general} source is defined as an infinite sequence
\[
{\bf X} = \left\{ X^n = \left( X_1^{(n)}X_2^{(n)}\cdots X_n^{(n)}  \right) \right\}_{n=1}^\infty
\]
of $n$-dimensional random variables $X^n$, where each component random variable $X_i^{(n)}$ takes values in a {\it countable} set ${\cal X}$.
It should be noted that each component of $X^n$ may change depending on block length $n$.
This implies that even {\it consistency} condition, which means that for any integers $m$, $n$ $(m < n)$, $X^{(m)}_i = X^{(n)}_i$ holds, may not hold. 

Variable-length codes are characterized as follows.
Let
\[
\varphi_n : {\cal X}^n \rightarrow {\cal U}^{\ast}, \ \  \psi_n : \{\varphi_n({\bf x})\}_{{\bf x} \in {\cal X}^n} \rightarrow {\cal X}^n,
\]
be a variable-length encoder and a decoder, respectively, where ${\cal U} = \{ 1,2,\cdots,K\}$ 
is called the code alphabet and ${\cal U}^{\ast}$ is the set of all finite-length strings over ${\cal U}$ excluding the null string.

We consider the situation that there are {\it unequal costs} on code symbols. Let us define the cost function over ${{\cal U}}$ considered in this paper.
Each code symbol $u \in {\cal U}$ is assigned the corresponding cost $c(u)$ such that $0 < c(u) < \infty$, and the additive cost $c({\bf u})$ of ${\bf u} = u_1, u_2, \cdots u_k \in {\cal U}^k$ is defined by
\[
c({\bf u}) = \sum_{i=i}^k c(u_i).
\]
In particular, we denote 
$c_{max} = \max_{u \in {\cal U}}c(u)$ for short.
This cost function is called memoryless cost function. A generalization of this cost function is discussed in Section VII.

We only consider variable-length codes satisfying prefix condition.
It should be noted that every variable-length code with prefix condition over {\it unequal costs},  satisfies
\begin{equation} \label{kraft}
\sum_{{\bf x}\in{\cal X}^n}K^{-\alpha_c c(\varphi_n({\bf x}))} \leq 1,
\end{equation}
where $\alpha_c$ is called {\it cost capacity} and defined as the positive unique root $\alpha$ of the equation \cite{CK}:
\begin{equation} \label{capacity}
\sum_{u \in {\cal U}} K^{-\alpha c(u)} =1.
\end{equation}
Throughout this paper, the logarithm is taken to the base $K$.
\subsection{Overflow Probability of Codeword Cost}
The overflow probability of codeword length is defined as follows:
\begin{definition}{\cite{Merhav91:overflow}}
Given a threshold $R$, the overflow probability of variable-length encoder $\varphi_n$ is defined by
\begin{equation}  \label{eq:2-1-0}
\varepsilon_n(\varphi_n, R) = \Pr\left\{ \frac{1}{n}l(\varphi({X^n})) >R \right\},
\end{equation}
where $l()$ denotes the length function.
\end{definition}

In this paper, we generalize the above overflow probability not only to the case for {\it unequal costs }on code symbols but also for finer evaluation of the overflow probability.
To this end, we consider the overflow probability of codeword cost as follows:
\begin{definition}[Overflow Probability of Codeword Cost]
Given some sequence $\{\eta_n\}_{n=1}^\infty$, where $0 < \eta_n < \infty$ for each $n=1,2,\cdots$, the overflow probability of variable-length encoder $\varphi_n$ is defined by
\begin{equation}\label{gof}
\varepsilon_n(\varphi_n, \eta_n) = \Pr\left\{ c(\varphi({X^n})) > \eta_n \right\}.
\end{equation}
\end{definition}
\begin{remark}
Nomura and Matsushima \cite{Nomura2011} have considered the overflow probability with respect to the codeword length, that is, $ \Pr\left\{ l(\varphi({X^n})) > \eta_n \right\}$ and derived the achievability of the first-order and the second-order sense.
Kosut and Sankar \cite{KS2013} have also defined the similar probability in the case of codeword length and derived the upper bound in universal setting.
\end{remark}

Since $\{\eta_n\}_{n=1}^\infty$ is an arbitrary sequence, the above definition is general.
In particular, we shall consider the following two types of overflow probability in this paper:
\begin{enumerate}
\item $\eta_n = nR$,
\item $\eta_n = na + \sqrt{n}L.$
\end{enumerate}
\begin{remark} \label{remark0}
If we set $ \eta_n = nR$ for all $n=1,2,\cdots$, that is, in the first case, the overflow probability can be written as
\begin{eqnarray*}
\varepsilon_n(\varphi_n,nR) & = & \Pr\left\{ c(\varphi_n(X^n)) > nR \right\} = \Pr\left\{ \frac{1}{n} c(\varphi_n(X^n)) > R \right\}.
\end{eqnarray*}
Thus, in the case that $\eta_n = nR$, the overflow probability defined by (\ref{gof}) means the probability that the codeword cost per symbol exceeds some constant $R$.
This is a natural extension of the overflow probability of codeword length to the overflow probability of codeword cost defined by (\ref{eq:2-1-0}).

On the other hand, in the analysis of fixed-length coding problem, Hayashi \cite{Hayashi} has shown the {\it second-order} asymptotics, which enables us a finer evaluation of achievable rate. A coding theorem from the view point of the {\it second-order} asymptotics have been also analyzed by Kontoyiannis \cite{Kon}.  
Analogously to their results, we evaluate the overflow probability in the {\it second-order} sense.
To do so, we consider the second case: $\eta_n = na+L\sqrt{n}$ for all $n=1,2,\cdots$.
Hereafter, if we consider the overflow probability in the case $\eta_n = na+L\sqrt{n}$, we call it the {\it second-order} overflow probability given $a$ in this paper, while in the first case it is called the {\it first-order} overflow probability.
The {\it second-order} overflow probability given $a$ of variable-length encoder $\varphi_n$ with threshold $L$ is written as
\begin{eqnarray*}
\varepsilon_n\left(\varphi_n,na+L\sqrt{n}\right) & = & \Pr\left\{ c(\varphi_n(X^n)) >  n a + L\sqrt{n} \right\} = \Pr\left\{\frac{ c(\varphi_n(X^n)) -  n a }{\sqrt{n}}  > L \right\}.
\end{eqnarray*}

It should be noted that since we assume that $\eta_n$ satisfies $0 < \eta_n < \infty$,  $0<R<\infty$ must hold, while $L$ can be negative number. 
\end{remark}

Throughout in this paper, we are interesting in the following achievability:
\begin{definition}[$\varepsilon$-achievable overflow thresholds] \label{def:overflow}
Sequence $\{\eta_n \}_{n=1}^\infty$ is called a sequence of $\varepsilon$-achievable overflow thresholds for the source if there exist a variable-length encoder ${\varphi_n}$ such that
\[
\limsup_{n \to \infty} \varepsilon_n(\varphi_n, \eta_n) \leq \varepsilon.
\]
\end{definition}
%
%
%
%
%
%
\section{Relationship between the overflow probability of variable-length coding and the error probability of fixed-length coding}
Uchida and Han \cite{Uchida01:overflow}, and Nomura, Matsushima and Hirasawa \cite{Nomura2007a} have derived the relationship between the overflow probability of variable-length coding and the error probability of fixed-length coding in the meaning of codeword length.
Analogously, we first reveal a deep relationship between the variable-length coding with codeword cost and the fixed-length coding.

Let $\varphi_n^f: {\cal X}^n \to {\cal U}_{M_n}$, $\psi_n^f: {\cal U}_{M_n} \to {\cal X}^n$ be a fixed-length encoder and a decoder, respectively, for source ${\bf X} = \{ X^n \}_{n=1}^\infty$, where $\ {\cal U}_{M_n} \stackrel{\mathrm{def}}{=} \{1,2, \cdots, M_n \} $ denotes a code set of size $M_n$. 
The decoding error probability $\varepsilon_n$ is given by 
$
\varepsilon^f_n \stackrel{\mathrm{def}}{=} \Pr \left\{ X^n \neq \psi_n^f \left(\varphi^f_n(X^n) \right) \right\}.
$
Such a code is denoted by $\left(n, M_n, \varepsilon^f_n \right)$.

We then define the $\varepsilon$-achievability in the fixed-length codes that is analogous to Definition \ref{def:overflow}.

\begin{definition} \label{def:ffachievable}
Sequence $\{\eta_n \}_{n=1}^\infty$ is called a sequence of $\varepsilon$-achievable fixed-length for the source if there exist an $\left(n, M_n, \varepsilon^f_n\right)$ code such that
\[
\limsup_{n \to \infty} \varepsilon^f_n \leq \varepsilon, \quad \limsup_{n \to \infty} \left( \log M_n - \eta_n \right) \leq 0.
\]
\end{definition}
The following theorem reveals the relationship between the overflow probability of variable-length coding and the error probability of fixed-length coding.
\begin{theorem}[Equivalence Theorem] \label{theo:0-1} \ 
\begin{enumerate}
\item Assuming that $\{\eta_n\}_{n=1}^\infty$ is a sequence of $\varepsilon$-achievable fixed-length, then $\left\{\frac{1}{\alpha_c} \eta_n + c\right\}_{n=1}^\infty$ is a sequence of $\varepsilon$-achievable overflow thresholds, where $c$ denotes the constant term which depends on the cost function and the source.
\item Assuming that $\{\eta_n\}_{n=1}^\infty$ is a sequence of $\varepsilon$-achievable overflow thresholds, then $\left\{{\alpha_c} \eta_n \right\}_{n=1}^\infty$ is a sequence of $\varepsilon$-achievable fixed-length.
\end{enumerate}
\end{theorem}
\begin{remark}
In \cite{Nomura2007a}, the equivalence theorem between the overflow probability of variable-length coding with codeword {\it length} and the error probability of fixed-length coding has already been established. This theorem can be considered as a generalization of their result to the variable-length codes with codeword {\it cost}.
\end{remark}
\begin{remark}
It should be emphasized that the proof of this theorem is substantially the same with the proof of \cite[Theorem 5]{Uchida01:overflow}.
\end{remark}

\begin{IEEEproof} The proof consists of two parts.

1) We first prove the first statement.
Suppose that $\{\eta_n\}_{n=1}^\infty$ is a sequence of $\varepsilon$-achievable fixed-length, then there exists an $\left(n, M_n, \varepsilon^f_n \right)$ code such that
\begin{equation} \label{eq:ff1}
\limsup_{n \to \infty} \varepsilon^f_n \leq \varepsilon,
\end{equation}
\begin{equation} \label{eq:ff2}
\limsup_{n \to \infty} \left( \log M_n - \eta_n \right) \leq 0.
\end{equation}
By using this $\left(n, M_n, \varepsilon^f_n \right)$ code, we define the set $T_n$ as
\[
T_n = \left\{ {\bf x} \in {\cal X}^n \left| {\bf x} = \psi_n^{f{\ast}}\left(\varphi^{f{\ast}}_n({\bf x}) \right) \right. \right\},
\]
where $\left(  \psi_n^{f{\ast}},\varphi^{f{\ast}} \right)$ denotes the pair of coding and decoding function satisfying (\ref{eq:ff1}) and (\ref{eq:ff2}).
By using this set $T_n$, we construct the variable-length encoder $\varphi_n$ as follows:
\[
\varphi_n({\bf x}) =  \left\{ \begin{array}{cc}
0 \ast \varphi_n^{(1)}({\bf x}) & \mbox{if } {\bf x} \in T_n \\
1 \ast \varphi_n^{(2)}({\bf x}) & \mbox{if } {\bf x} \in T_n^c , \\
\end{array} \right.
\]
where, $\varphi_n^{(1)}: T_n \to {\cal U}^\ast$ denotes the encoding function proposed by Han and Uchida \cite{Han00:cost},
for random variable $Z_n$ distributed in uniformly on $T_n$ and $\varphi_n^{(2)}: {\cal X}^n \setminus T_n \to {\cal U}^\ast$ is an arbitrary variable-length encoder.
Notice here that from the property of the code, it holds that
\begin{eqnarray} \label{length}
c(\varphi_n^{(1)}({\bf x})) \leq - \frac{1}{\alpha_c} \log P_{X^n}({\bf x}) + \frac{\log 2}{\alpha_c} + c_{max},
\end{eqnarray}
for all $n=1,2,\cdots$.

Now, we evaluate the overflow probability of this code.
From (\ref{length}), we have for ${\bf x} \in T_n$
\begin{eqnarray*}
c(\varphi_n({\bf x})) & \leq & \frac{1}{\alpha_c} \log |T_n| + \frac{\log 2}{\alpha_c} + c_{max} + c(0) \\
& \leq &  \frac{1}{\alpha_c} \log M_n+ \frac{\log 2}{\alpha_c} + 2c_{max}.
\end{eqnarray*}

On the other hand, from (\ref{eq:ff2}) we have, for $\forall \delta >0$
\[
\log M_n < \eta_n + \delta,
\]
for sufficiently large $n$.
Thus, from the construction of this variable-length code, for $\forall{\bf x} \in T_n$, the codeword cost is upper bounded by
\begin{eqnarray*}
c(\varphi_n({\bf x})) & < & \frac{1}{\alpha_c} \left( \eta_n + \delta \right)  +  \frac{\log 2}{\alpha_c} + 2c_{max} \\
& = & \frac{1}{\alpha_c} \eta_n +  \frac{\log 2}{\alpha_c} + 2c_{max} + \frac{\delta}{\alpha_c},
\end{eqnarray*}
for sufficiently large $n$.

Setting $c = \frac{\log 2}{\alpha_c} + 2c_{max} $,
this means that 
\begin{eqnarray*}
\lefteqn{\varepsilon_n \left(\varphi_n, \frac{1}{\alpha_c}\eta_n + c + \frac{\delta}{\alpha_c}  \right)} \\
 & \leq & \Pr\left\{c(\varphi_n(X^n)) \geq  \frac{1}{\alpha_c} \eta_n + c + \frac{\delta}{\alpha_c} \right\} \\
& \leq & \Pr\left\{ X^n \in T_n^c \right\},
\end{eqnarray*}
for sufficiently large $n$.

Here, from (\ref{eq:ff1}) and the definition of $T_n$, we have
\begin{align*}
\lefteqn{\limsup_{n \to \infty} \varepsilon_n \left(\varphi_n, \frac{1}{\alpha_c}\eta_n + c + \frac{\delta}{\alpha_c}  \right)} \\
& \leq \limsup_{n \to \infty} \Pr\left\{ X^n \in T_n^c \right\} \\
& \leq \varepsilon.
\end{align*}
Noting that $\delta$ is arbitrarily small, this means that the first statement holds.
%
%
%
%

2) Suppose that $\{\eta_n \}_{n=1}^\infty$ is a sequence of $\varepsilon$-achievable overflow thresholds. Then, there exists a variable-length code satisfying
\begin{equation} \label{eq:2-1-0-1}
\limsup_{n \to \infty} \varepsilon_n(\varphi_n, \eta_n) \leq \varepsilon.
\end{equation}
By using this variable-length code
we define a set $T_n$ as
\[
T_n = \left\{ {\bf x} \in {\cal X}^n \left| c(\varphi^{\ast}_n({\bf x})) \leq \eta_n \right. \right\},
\]
where $\varphi^{\ast}$ denotes the variable-length encoder satisfying (\ref{eq:2-1-0-1}).

Here, from (\ref{kraft}) it holds that
\begin{eqnarray*}
1 & \geq & \sum_{{\bf x} \in {\cal X}^n} K^{-\alpha_c c(\varphi^\ast_n({\bf x}))} \geq \sum_{{\bf x} \in {T_n}} K^{-\alpha_c c(\varphi^\ast_n({\bf x}))} \nonumber \\
& \geq & |T_n| K^{-\alpha_c \eta_n}.  
\end{eqnarray*}
Thus, we have 
\begin{equation} \label{eq:ff3}
|T_n| \leq K^{\alpha_c \eta_n}.
\end{equation}

Now, we define the fixed-length encoder with $M_n = |T_n|$ as
\[
\varphi_n^f({\bf x}) =  \left\{ \begin{array}{cc}
1,2,\dots,M_n & \mbox{if } {\bf x} \in T_n \\
1 & \mbox{if } {\bf x} \in T_n^c , \\
\end{array} \right.
\]
and the decoder $\psi_n^f(i): {\cal U}_{M_n} \to T_n$ as the mapping such that
$\psi_n^f(i) = {\bf x}$ if $\varphi_n^f({\bf x}) = i$ for some ${\bf x} \in T_n$.
%
Then, from (\ref{eq:ff3}) and the fact that $M_n = |T_n|$ we have
\[
\limsup_{n \to \infty} \left( \log M_n - \alpha_c \eta_n \right) \leq 0.
\]

On the other hand since  $\{\eta_n \}_{n=1}^\infty$ is a sequence of $\varepsilon$-achievable overflow thresholds, we have
\[
\limsup_{n \to \infty} \Pr \left\{ {\bf x} \in T_n^c \right\} \leq \varepsilon.
\]
Here, noting that the error probability of this fixed-length code is given by $\varepsilon^f_n = \Pr \left\{ {\bf x} \in T_n^c \right\}$, the second statement has been proved.
\end{IEEEproof}

The definition of the $\varepsilon$-achievability for fixed-length codes (Definition \ref{def:ffachievable}) is very general and it includes the ordinary first-order and the second-order achievability.
Hence from Theorem \ref{theo:0-1} and the previous results for fixed-length codes such as \cite{Hayashi} and \cite{Han}, we can obtain analogous results for the overflow probability of variable-length codes (see,  Remark \ref{remark:5-1} in Section V and Theorem \ref{theo:4-1} in Section VI for example).
In the following section, however, we derive several theorems from another information-spectrum approach in order to see the logic underlying the whole process of variable-length codes with codeword cost.

%
%
%
%
%
%
%
\section{First-order and Second-order achievability}
Hereafter, we consider the first-order and the second-order achievable threshold as described in Remark \ref{remark0}.
In the first-order case, we are interested in the infimum of threshold $R$ that we can achieve.
This is formalized as follows.
\begin{definition} 
Given $0 \leq \varepsilon < 1 $, $R$ is called an $\varepsilon$-achievable overflow threshold for the source if there exists a variable-length encoder $\varphi_n$ such that
\begin{equation*}
\limsup_{n \rightarrow \infty} \varepsilon_n(\varphi_n,nR) \leq \varepsilon.
\end{equation*}
\end{definition}
\begin{definition} [Infimum of $\varepsilon$-achievable overflow threshold]
\[
R(\varepsilon|{\bf X}) \stackrel{\mathrm{def}}{=} \inf \left\{ R| R \mbox{ is an $\varepsilon$-achievable overflow threshold} \right\}.
\]
\end{definition}

Also, in the analysis of second-order overflow probability, we define the achievability:
\begin{definition}
Given $0 \leq \varepsilon < 1 $ and $0 < a < \infty$, $L$ is called an $(\varepsilon,a)$-achievable overflow threshold for the source, if there exists a variable-length encoder $\varphi_n$ such that
\begin{equation*}
\limsup_{n \rightarrow \infty} \varepsilon_n\left(\varphi_n,na+L\sqrt{n}\right) \leq \varepsilon.
\end{equation*}
\end{definition}
\begin{definition}[Infimum of $(\varepsilon,a)$-achievable overflow threshold]
\begin{equation*}
L(\varepsilon,a|{\bf X}) \stackrel{\mathrm{def}}{=} \inf \left\{ L| L \mbox{ is an $(\varepsilon,a)$-achievable overflow threshold} \right\}.
\end{equation*}
\end{definition}

As described in the previous section, we demonstrate the infimum of $\varepsilon$-achievable overflow threshold and $(\varepsilon,a)$-achievable overflow threshold not via Theorem \ref{theo:0-1} but via another {\it information-spectrum} approach.
To do so, we show two lemmas that have important roles to derive theorems. 
\begin{lemma} \label{lemma1}
For any general sources ${\bf X}$ and any sequence of positive number $\{\eta_n \}_{n=1}^\infty$, there exists a variable-length encoder $\varphi_n$ that satisfies
\begin{eqnarray}
\varepsilon_n(\varphi_n,\eta_n) < \Pr\left\{ z_n P_{X^n}(X^n) \leq K^{-\alpha_c \eta_n} \right\} + z_n K^{\alpha_c c_{max}+1},
\end{eqnarray}
for $n=1,2,\cdots$, where $\{ z_n \}_{n=1}^\infty$ is a given sequence of an arbitrary number satisfying $z_i > 0$ for $i=1,2,\cdots$ and $\alpha_c$ denotes the {\it cost capacity} defined in (\ref{capacity}).
\end{lemma}
\begin{IEEEproof}
 We use the code proposed by Han and Uchida \cite{Han00:cost}.
Then, from the property of the code, it holds that
\begin{eqnarray} \label{length}
c(\varphi_n^{\ast}({\bf x})) \leq - \frac{1}{\alpha_c} \log P_{X^n}({\bf x}) + \frac{\log 2}{\alpha_c} + c_{max},
\end{eqnarray}
for all $n=1,2,\cdots$, where $\varphi_n^{\ast}$ denotes the encoder of the code.
Furthermore, we set the decoder as the inverse mapping of $\varphi_n^{\ast}$ that is, $\psi_n = \varphi_n^{\ast -1}$. 
Please note that the code is a uniquely decodable variable-length code for {\it general} sources with countably infinite source alphabet.

Next, we shall evaluate the overflow probability of this code.
Set 
\[
A_n = \left\{{\bf x}\in {\cal X}^n \left|  z_n P_{X^n}(X^n) \leq K^{-\alpha_c \eta_n} \right. \right\},
\]
\begin{equation*} 
S_n = \left\{ {\bf x} \in {\cal X}^n \left|  c (\varphi_n^{\ast}({\bf x})) > \eta_n \right. \right\}.
\end{equation*}
The overflow probability is given by
\begin{eqnarray} \label{eq:1-1}
\varepsilon_n(\varphi_n, \eta_n) & = & \Pr \left\{ X^n \in S_n \right\} = \sum_{{\bf x} \in S_n} P_{X^n}({\bf x}) \nonumber \\
& = & \sum_{{\bf x} \in S_n \cap A_n} P_{X^n}({\bf x}) + \sum_{{\bf x} \in S_n \cap A_n^c} P_{X^n}({\bf x}) \nonumber \\
& \leq & \Pr \left\{ X^n\!\in\!A_n \right\} + \sum_{{\bf x} \in S_n \cap A_n^c} P_{X^n}({\bf x}),
\end{eqnarray}
where $A^c$ denotes the complement set of the set $A$.

Since (\ref{length}) holds, for $\forall {\bf x}\in S_n$, we have
\begin{eqnarray*}
 - \frac{1}{\alpha_c} \log P_{X^n}({\bf x}) + \frac{\log 2}{\alpha_c} +c_{max}  > \eta_n.
\end{eqnarray*}
Thus, we have
\begin{eqnarray*}
 P_{X^n}({\bf x}) < K^{- \alpha_c \left(\eta_n - c_{max}\right) + \log 2},
\end{eqnarray*}
for $\forall {\bf x} \in S_n$.
Substituting the above inequality into (\ref{eq:1-1}), we have
\begin{eqnarray} \label{eq:1-2}
\varepsilon_n(\varphi_n, \eta_n) & < & \Pr \left\{ X^n \in A_n \right\} + \sum_{{\bf x} \in S_n \cap A_n^c} K^{- \alpha_c \left(\eta_n - c_{max}\right)+ \log 2} \nonumber \\
& = & \Pr \left\{ X^n \in A_n \right\} + \left| S_n \cap A_n^c \right|  K^{- \alpha_c \left(\eta_n - c_{max}\right)+ \log2}.
\end{eqnarray}
Here, from the definition of $A_n$, for $\forall {\bf x} \in A_n^c$, it holds that
\begin{equation*}
P_{X^n}(X^n) > \frac{K^{ - \alpha_c \eta_n}}{z_n}.
\end{equation*}
Thus, we have
\begin{eqnarray*}
1\!&\geq\!&\sum_{{\bf x} \in {\cal X}^n} P_{X^n}(X^n) \geq \sum_{{\bf x} \in A_n^c} P_{X^n}(X^n) > \sum_{{\bf x} \in A_n^c}  \frac{K^{ - \alpha_c \eta_n}}{z_n} = \left| A_n^c \right|  \frac{K^{ - \alpha_c \eta_n}}{z_n} .
\end{eqnarray*}
This means that
\begin{equation} \label{eq:1-3}
\left| S_n \cap A_n^c \right| \leq  \left| A_n^c \right| < z_n K^{ \alpha_c \eta_n}.
\end{equation}
Substituting (\ref{eq:1-3}) into (\ref{eq:1-2}), we have
\begin{eqnarray*}
\varepsilon_n(\varphi_n, \eta_n) & < & \Pr \left\{ X^n \in A_n \right\} + z_n K^{\alpha_c \eta_n} K^{- \alpha_c \left(\eta_n - c_{max}\right)+ \log 2} \nonumber \\
& \leq & \Pr \left\{ X^n\in A_n \right\} + z_n K^{\alpha_c c_{max}+1},
\end{eqnarray*}
since $\log 2 \leq 1$.
Therefore, we have proved the lemma. 
\end{IEEEproof}
%
%
%
%
%
\begin{lemma} \label{lemma2}
For any variable-length code and any sequence $\{\eta_n \}_{n=1}^\infty$, it holds that
\begin{eqnarray*}
\varepsilon_n(\varphi_n,\eta_n) \geq \Pr\left\{ P_{X^n}(X^n) \leq z_n K^{- \alpha_c \eta_n} \right\} - z_n,
\end{eqnarray*}
for $n=1,2,\cdots$, where $\{ z_n \}_{n=1}^\infty$ is a given sequence of an arbitrary number satisfying $z_i > 0$ for $i=1,2,\cdots$.
\end{lemma}
\begin{IEEEproof}
Let $\varphi_n$ be an encoder and decoder of variable-length code.
Set 
\[
B_n = \left\{ {\bf x} \in {\cal X}^n \left| P_{X^n}(X^n) \leq z_n K^{-\alpha_c \eta_n} \right. \right\}.
\]
\begin{equation} \label{sn}
S_n = \left\{ {\bf x} \in {\cal X}^n \left|  c (\varphi_n({\bf x})) > \eta_n \right. \right\}.
\end{equation}

Then, by using $S_n$ defined by (\ref{sn}) we have
\begin{align}
\Pr \left\{ P_{X^n}(X^n) \leq z_n K^{-\alpha_c \eta_n} \right\} &= \sum_{{\bf x} \in B_n} P_{X^n}({\bf x})  = \sum_{{\bf x} \in B_n \cap S_n} P_{X^n}({\bf x}) + \sum_{{\bf x} \in B_n \cap S_n^c } P_{X^n}({\bf x}) \nonumber \\
& \leq \sum_{{\bf x} \in  S_n} P_{X^n}({\bf x}) + \sum_{{\bf x} \in B_n \cap S_n^c } P_{X^n}({\bf x}) \nonumber \\
& \leq \varepsilon_n(\varphi_n,\eta_n) + \sum_{{\bf x} \in B_n \cap S_n^c } P_{X^n}({\bf x}).
\end{align}
On the other hand, for $\forall {\bf x} \in B_n$ it holds that
\begin{equation*}
P_{X^n}({\bf x}) \leq z_n K^{-\alpha_c \eta_n}.
\end{equation*}
Thus, we have 
\begin{align} \label{eq:2-2}
\Pr \left\{ P_{X^n}(X^n) \leq z_n K^{-\alpha \eta_n} \right\}  &\leq  \varepsilon_n(\varphi_n,\eta_n) + \sum_{{\bf x} \in B_n \cap S_n^c } P_{X^n}({\bf x}) \nonumber \\
& \leq  \varepsilon_n(\varphi_n,\eta_n) + \sum_{{\bf x} \in B_n \cap S_n^c } z_n K^{-\alpha_c \eta_n} \nonumber \\
& =  \varepsilon_n(\varphi_n,\eta_n) + \left| B_n \cap S_n^c \right| z_n K^{-\alpha_c \eta_n}.
\end{align}
Here, from (\ref{kraft}), we have
\begin{align*}
1 & \geq  \sum_{{\bf x} \in {\cal X}^n}  K^{ - \alpha_c c(\varphi_x({\bf x})) }  \geq  \sum_{{\bf x} \in S_n^c}  K^{ - \alpha_c c(\varphi_x({\bf x})) } \geq  \sum_{{\bf x} \in S_n^c}  K^{ - \alpha_c \eta_n } = \left| S_n^c \right| K^{ - \alpha_c \eta_n} .
\end{align*}
This mean that
\begin{equation} \label{eq:2-2-1}
\left| B_n \cap S_n^c \right| \leq \left| S_n^c \right| \leq K^{ \alpha_c \eta_n}
\end{equation}

Hence, substituting (\ref{eq:2-2-1}) into (\ref{eq:2-2}), we have
\begin{align*}
\Pr \left\{ P_{X^n}(X^n) \leq a_n K^{ - \eta_n} \right\}
& \leq  \varepsilon_n(\varphi_n,\eta_n) + \left| B_n \cap S_n^c \right| z_n K^{-\alpha_c \eta_n} \\
& \leq  \varepsilon_n(\varphi_n,\eta_n) +  K^{\alpha_c{\eta_n} } z_n K^{-\alpha_c \eta_n} \\
& =  \varepsilon_n(\varphi_n,\eta_n) +  z_n .
\end{align*}
Therefore, we have proved the lemma.
\end{IEEEproof}
%
%
%
%
%
\section{Infimum of $\varepsilon$-achievable overflow threshold}
\subsection{General Formula for the Infimum of $\varepsilon$-achievable Overflow Threshold}
In this subsection, we determine $R(\varepsilon|{\bf X})$ for {\it general} sources.
Before showing the theorem, we define the function $F(R)$ as follows:
\[
F(R) \stackrel{\mathrm{def}}{=} \limsup_{n \to \infty} \Pr \left\{ \frac{1}{n\alpha_c} \log \frac{1}{P_{X^n}(X^n)} \geq R \right\}
\]
The following theorem is one of our main results:
\begin{theorem} \label{theorem1} For $0 \leq \forall \varepsilon < 1$, it holds that 
\begin{eqnarray*}
R(\varepsilon|{\bf X}) = \inf \left\{ R \left| F(R) \leq \varepsilon \right. \right\}.
\end{eqnarray*}
\end{theorem}
\begin{IEEEproof}
The proof consists of two parts. \\
(Direct Part)
Let $R_0$ be as
\begin{equation*}
R_0 = \inf \left\{ R \left| F(R) \leq \varepsilon \right. \right\},
\end{equation*}
for short.
Then, in this part we show that
\begin{equation} \label{eq:4-1}
R(\varepsilon|{\bf X}) \leq R_0 + \gamma,
\end{equation}
for any $\gamma >0$ by showing that $R_0$ is an $\varepsilon$-achievable overflow threshold for the source.
Let $\eta_n$ be as $\eta_n = n(R_0+ \gamma)$, then from Lemma \ref{lemma1} there exists a variable-length code $(\varphi_n,\psi_n)$ that satisfies
\begin{eqnarray*}
\varepsilon_n(\varphi_n,n(R_0+\gamma)) < \Pr\left\{ z_n P_{X^n}(X^n) \leq K^{ - n\alpha_c( R_0+\gamma)} \right\} + z_n K^{\alpha_c c_{max}+1},
\end{eqnarray*}
for $n=1,2,\cdots$. 
Thus, we have
\begin{align*}
\varepsilon_n(\varphi_n,n(R_0+\gamma))& < \Pr\left\{ z_n P_{X^n}(X^n) \leq K^{ - n\alpha_c( R_0+\gamma)} \right\} + z_n K^{\alpha_c c_{max}+1} \nonumber \\
& =  \Pr\left\{ \frac{1}{n} \log \frac{1}{z_n P_{X^n}(X^n)} \geq \alpha_c (R_0+\gamma) \right\} + z_n K^{\alpha_c c_{max}+1} \nonumber \\
& =  \Pr\left\{ \frac{1}{n \alpha_c} \log \frac{1}{P_{X^n}(X^n)} \geq R_0 +\gamma + \frac{1}{n\alpha_c} \log z_n \right\} + z_n K^{\alpha_c c_{max}+1},
\end{align*}
for $n=1,2,\cdots$.

Notice here that $z_n > 0$ is an arbitrary number. Setting $z_n = K^{-\sqrt{n} \gamma}$, we have
\begin{align*}
\varepsilon_n(\varphi_n,n(R_0+\gamma)) & < \Pr\left\{ \frac{1}{n \alpha_c} \log \frac{1}{ P_{X^n}(X^n)}\geq R_0 \!+\! \gamma \!-\! \frac{\sqrt{n} \gamma}{n\alpha_c } \right\}  +K^{-\sqrt{n}\gamma + \alpha_c c_{max}+1} \\
& < \Pr\left\{ \frac{1}{n \alpha_c} \log \frac{1}{ P_{X^n}(X^n)}\geq R_0 + \frac{\gamma}{2} \right\} +K^{-\sqrt{n}\gamma + \alpha_c c_{max}+1} \\
& < \Pr\left\{ \frac{1}{n \alpha_c} \log \frac{1}{ P_{X^n}(X^n)}\geq R_0 \right\} +K^{-\sqrt{n}\gamma + \alpha_c c_{max}+1},
\end{align*}
for sufficiently large $n$, because $\frac{\gamma}{2}> \frac{ \gamma}{\sqrt{n} \alpha_c}$ as $n \to \infty$.
Thus, since $\alpha_c$ and $c_{max}$ are positive constants, by taking $\limsup_{n \to \infty}$, we have
\begin{align*}
\limsup_{n \to \infty} \varepsilon_n(\varphi_n,n(R_0+\gamma)) \leq  \limsup_{n \to \infty} \Pr\left\{ \frac{1}{n \alpha_c} \log \frac{1}{ P_{X^n}(X^n)} \!\geq\! R_0  \right\}.
\end{align*}

Hence, from the definition of $R_0$ we have
\begin{IEEEeqnarray*}{rCl}
\limsup_{n \to \infty} \varepsilon_n(\varphi_n,n(R_0+\gamma)) & \leq & \varepsilon.
\end{IEEEeqnarray*}
Noting that $\gamma >0$ is arbitrarily small, the direct part has been proved. \\
%
%
(Converse Part) \\
Assuming that $R_1$ satisfying
\begin{equation} \label{eq:4-2-0}
R_1 < \inf \left\{ R \left| F(R) \leq \varepsilon \right. \right\},
\end{equation}
is an $\varepsilon$-achievable overflow threshold, then we shall show a contradiction.

Let $\eta_n$ be as $\eta_n = nR_1$. Then from Lemma \ref{lemma2} for any sequence $\{ z_n \}_{n=1}^\infty$ ( $z_i > 0$ $i=1,2,\cdots$) and any variable-length code we have
\begin{eqnarray*}
\varepsilon_n(\varphi_n, nR_1) > \Pr\left\{ P_{X^n}(X^n) \leq z_n K^{ - n\alpha_cR_1} \right\} - z_n,
\end{eqnarray*}
for $n=1,2,\cdots$.
Thus, we have for any variable-length code 
\begin{align*}
\varepsilon_n(\varphi_n,nR_1) 
&> \Pr\left\{ P_{X^n}(X^n) \leq z_n K^{ - n\alpha_cR_1} \right\} - z_n \\
&= \Pr\left\{ \frac{1}{n\alpha_c} \log \frac{1}{P_{X^n}(X^n)} \geq R_1 - \frac{1}{n\alpha_c}\log z_n \right\} - z_n .
\end{align*}
Set $z_n = K^{-n \gamma}$, where $\gamma >0$ is a small constant that satisfies
\begin{equation} \label{eq:4-2-1}
R_1 + \frac{\gamma}{\alpha_c} < \inf \left\{ R \left| F(R) \leq \varepsilon \right. \right\}.
\end{equation}
Since we assume that (\ref{eq:4-2-0}) holds, it is obvious that there exists $\gamma >0$ that satisfies the above inequality.
Then, we have
\begin{align*}
\varepsilon_n(\varphi_n,nR_1)& >  \Pr\left\{ \frac{1}{n\alpha_c} \log \frac{1}{P_{X^n}(X^n)} \geq R_1 + \frac{\gamma}{\alpha_c} \right\} - K ^{ -n\gamma}.
\end{align*}
Hence, we have
\begin{align*}
\limsup_{n \to \infty}\varepsilon_n(\varphi_n,nR_1) & \geq  \limsup_{n \to \infty} \Pr\left\{ \frac{1}{n} \log \frac{1}{P_{X^n}(X^n)} \geq R_1 + \frac{\gamma}{\alpha_c} \right\}  > \varepsilon,
\end{align*}
where the last inequality is derived from (\ref{eq:4-2-1}) and the definition of $F(R)$.

On the other hand, since we assume that $R_1$ is an $\varepsilon$-achievable overflow threshold, it holds that
\begin{equation*}
\limsup_{n \to \infty} \varepsilon_n\left(\varphi_n,nR_1 \right) \leq \varepsilon.
\end{equation*}
This is a contradiction.
Therefore the proof of converse part has been completed. 
\end{IEEEproof}
\begin{remark}\label{remark:5-1}
Theorem \ref{theorem1} also can be proved from Theorem \ref{theo:0-1} combined with the result of \cite{Steinberg} (see also, \cite[Theorem 1.6.1]{Han}) which shows the infimum of first-order $\varepsilon$-achievable fixed-length coding rate.
Theorem \ref{theorem2} in Section VI can also be derived by using Theorem \ref{theo:0-1} combined with \cite[Theorem 3]{Hayashi}.
\end{remark}

From the above theorem, we can show a corollary. Before describing the corollary, we define the spectral sup-entropy rate \cite{Han}\footnote{%
For any sequence $\{Z_n \}_{n=1}^{\infty}$ of real-valued random variables, we define the limit superior in probability of $\{Z_n \}_{n=1}^{\infty}$ by
$\mbox{p-}\limsup_{n \to \infty} Z_n = \inf \left\{ \beta | \lim_{n \to \infty} \Pr \{Z_n > \beta \} = 0 \right\}$ (cf.\cite{Han}) .
}
:
\begin{equation*}
\overline{H}({\bf X}) = \mbox{p-}\limsup_{n \to \infty} \frac{1}{n} \log \frac{1}{P_{X^n}(X^n)},
\end{equation*}
Then, the following corollary holds.
\begin{corollary} \label{coro1}
\begin{equation} \label{supentropy}
R(0|{\bf X}) = \frac{1}{\alpha_c}\overline{H}({\bf X}).
\end{equation} 
\end{corollary}
\subsection{Strong Converse Property}
Strong converse property is one of important properties for the source in fixed-length source coding problem \cite{Han}. When we consider the second-order achievability, we give appropriate first-order term. In many cases the first-order term is determined by considering the strong converse property and hence the strong converse property has an important meaning in the analysis of second-order achievability.

Analogously to the fixed-length coding problem, we can consider the strong converse property in the meaning of the overflow probability in variable-length codes.
In this subsection, we establish the strong converse theorem on the overflow probability of variable-length coding with codeword cost.
Let us begin with the definition of strong converse property treated in this paper.
\begin{definition}
Source ${\bf X}$ is said to satisfy the strong converse property, if any variable-length code $(\varphi_n,\psi_n)$ with the overflow probability $\varepsilon_n(\varphi_n, nR)$, where $R$ is an arbitrary rate satisfying $R < R(0|{\bf X})$, necessarily yields
\[
\lim_{n \to \infty} \varepsilon_n(\varphi_n, nR) = 1.
\]
\end{definition}

In order to state the strong converse theorem, we define the dual quantity of $\overline{H}({\bf X})$ as
\begin{equation*}
\underline{H}({\bf X}) = \mbox{p-}\liminf_{n \to \infty} \frac{1}{n} \log \frac{1}{P_{X^n}(X^n)},
\end{equation*}
which is called the spectral inf-entropy rate \cite{Han}\footnote{%
For any sequence $\{Z_n \}_{n=1}^{\infty}$ of real-valued random variables, we define the limit inferior in probability of $\{Z_n \}_{n=1}^{\infty}$ by
$\mbox{p-}\liminf_{n \to \infty} Z_n = \sup \left\{ \alpha | \lim_{n \to \infty} \Pr \{Z_n < \alpha \} = 0 \right\}$ (cf.\cite{Han}) .
}.
Then, we have the following theorem on the strong converse property.
\begin{theorem} \label{theo:scp}
Source ${\bf X}$ satisfies the strong converse property if and only if
\begin{eqnarray} \label{eq:scp}
\overline{H}({\bf X}) = \underline{H}({\bf X})
\end{eqnarray}
holds.
\end{theorem}
\begin{IEEEproof} This theorem can be proved by using the similar argument with the proof of \cite[Theorem 1.5.1]{Han}. For the completeness of this paper we give the proof in Appendix A.
\end{IEEEproof}

The theorem reveals that the strong converse property only depends on source {\bf X} and is independent on cost function.
\begin{remark} \label{remark:5-2}
For an i.i.d. source, the following relationship holds \cite{Han},
\[
H(X) = \overline{H}({\bf X}) = \underline{H}({\bf X}),
\]
where $H(X)$ denotes the entropy of the source.
Thus, any i.i.d. source satisfies the strong converse property.
This means that the infimum of $\varepsilon$-achievable overflow threshold $R(\varepsilon|{\bf X})$ is constant and is independent on $\varepsilon$.
\end{remark}
%
%
%
%
\section{Infimum of $(\varepsilon,a)$-achievable overflow threshold}
\subsection{General formula for the infimum of $(\varepsilon,a)$-achievable overflow threshold}
So far, we have considered the {\it first-order} achievable threshold. 
In this section, we consider the {\it second-order} achievability.
In the {\it second-order} case, the infimum $(\varepsilon,a)$-achievable overflow threshold for general sources is also determined by using Lemma \ref{lemma1} and Lemma \ref{lemma2}.

We define the function $F_a(R)$ given $a$ as follows, which is correspondence with the function $F(R)$ in {\it first-order} case.
\begin{equation*}
F_a(L) \stackrel{\mathrm{def}}{=} \limsup_{n \to \infty} \Pr \left\{ \frac{ - \log P_{X^n}(X^n) -n \alpha_c a }{\sqrt{n}\alpha_c} \geq  {L} \right\}.
\end{equation*}
Then, we have 
\begin{theorem} \label{theorem2} For $0 \leq \forall \varepsilon < 1$, it holds that 
\begin{eqnarray*}
L(\varepsilon,a|{\bf X}) = \inf \left\{ L \left| F_a(L) \leq \varepsilon \right. \right\}.
\end{eqnarray*}
\end{theorem}
\begin{IEEEproof}
The proof is similar to the proof of Theorem \ref{theorem1}. \\
(Direct Part)
Let $L_0$ be as
\begin{equation*}
L_0 = \inf \left\{ L \left| F_a(L) \leq \varepsilon \right. \right\},
\end{equation*}
for short.
Then, in this part we shall show that
\begin{equation} \label{eq:6-1}
L(\varepsilon,a|{\bf X}) \leq L_0 + \gamma,
\end{equation}
for any $\gamma >0$ by showing that $L_0$ is an $\varepsilon$-achievable overflow threshold for the source.
Let $\eta_n$ be as $\eta_n = n a + \sqrt{n}(L_0+\gamma)$, then from Lemma \ref{lemma1}, for any sequence $\{ z_n \}_{n=1}^\infty \ (z_i >0, i=1,2,\cdots)$ there exists a variable-length encoder $\varphi_n$ that satisfies
\begin{align*}
\varepsilon_n\left(\varphi_n, na + \sqrt{n}(L_0+\gamma) \right) < \Pr\left\{ z_n P_{X^n}(X^n) \leq K^{ - \alpha_c(na + \sqrt{n}L_0 + \sqrt{n} \gamma)} \right\} + z_n K^{\alpha_c c_{max} +1}.
\end{align*}
Thus, we have
\begin{align*}
\varepsilon_n\left(\varphi_n, na + \sqrt{n}(L_0+\gamma)\right)  < & \Pr\left\{ z_n P_{X^n}(X^n) \leq K^{ - \sqrt{n}\alpha_c(\sqrt{n}a + L_0 + \gamma) } \right\} + z_n K^{\alpha_c c_{max} +1} \nonumber \\
 = & \Pr\left\{ \frac{1}{\sqrt{n}\alpha_c} \log \frac{1}{z_n P_{X^n}(X^n)} \geq \sqrt{n}{a} + L_0 + \gamma \right\} + z_n K^{\alpha_c c_{max} +1} \nonumber \\
 = & \Pr\left\{ \frac{1}{\sqrt{n}\alpha_c} \log \frac{1}{ P_{X^n}(X^n)} \geq \sqrt{n}{a}\!+\!L_0\!+ \gamma + \!\frac{1}{\sqrt{n }\alpha_c}\log z_n \right\}  + z_n K^{\alpha_c c_{max} +1}.
\end{align*}
Let $z_n$ be as $z_n  = K^{-\sqrt[4]{n} \gamma}$, then we have
\begin{align*}
\varepsilon_n\left( \varphi_n, na + \sqrt{n}(L_0+\gamma)\right)  & <  \Pr\left\{ \frac{1}{\sqrt{n}\alpha_c} \log \frac{1}{ P_{X^n}(X^n)} \geq \sqrt{n}{a} + L_0 + \gamma -\!\frac{\sqrt[4]{n} \gamma}{\sqrt{n} \alpha_c} \right\} + K^{\alpha_c c_{max} +1} K^{-\sqrt[4]{n}\gamma} \\
& = \Pr\left\{ \frac{- \log P_{X^n}(X^n) - n\alpha_ca }{\sqrt{n}\alpha_c} \geq  L_0 +\gamma -\!\frac{ \gamma}{\sqrt[4]{n}\alpha_c}  \right\}+ K^{\alpha_c c_{max} +1-\sqrt[4]{n}\gamma} \\
& < \Pr\left\{ \frac{- \log P_{X^n}(X^n) - n\alpha_ca }{\sqrt{n}\alpha_c} \geq  L_0 + \frac{\gamma}{2}  \right\}+ K^{\alpha_c c_{max} +1-\sqrt[4]{n}\gamma} \\
& < \Pr\left\{ \frac{- \log P_{X^n}(X^n) - n\alpha_ca }{\sqrt{n}\alpha_c} \geq  L_0  \right\}+ K^{\alpha_c c_{max} +1-\sqrt[4]{n}\gamma}
\end{align*}
for sufficiently large $n$, because  $ \frac{\gamma}{2} > \!\frac{ \gamma}{\sqrt[4]{n}\alpha_c}$ holds for sufficiently large $n$.

By taking $\limsup_{n \to \infty}$, we have
\begin{align*}
\limsup_{n \to \infty} \varepsilon_n \left( \varphi_n,na + \sqrt{n}(L_0+\gamma) \right)  \leq  \limsup_{n \to \infty} \Pr\left\{ \frac{- \log P_{X^n}(X^n) - n\alpha_c a }{\sqrt{n}\alpha_c} \geq  L_0  \right\}.
\end{align*}
Hence, from the definition of $L_0$ we have
\begin{eqnarray*}
\limsup_{n \to \infty} \varepsilon_n\left( \varphi_n,na + \sqrt{n}(L_0+\gamma) \right) \leq \varepsilon.
\end{eqnarray*}
This means that (\ref{eq:6-1}) holds. 
Therefore, the direct part has been proved. \\
%
%
(Converse Part) \\
Assuming that $L_1$ satisfying
\begin{equation} \label{eq:6-2-0}
L_1 < \inf \left\{ L \left| F_a(L) \leq \varepsilon \right. \right\},
\end{equation}
is an $(\varepsilon,a)$-achievable second order overflow threshold, we shall show a contradiction.

From Lemma \ref{lemma2} for any sequence $\{ z_n \}_{n=1}^\infty \ (z_i >0, i=1,2,\cdots)$ and any variable-length encoder, it holds that
\begin{align*}
\varepsilon_n\left( \varphi_n,na + \sqrt{n}L_1 \right) \geq \Pr\left\{ P_{X^n}(X^n) \leq z_n K^{ - \alpha_c(na - \sqrt{n}L_1)} \right\} - z_n ,
\end{align*}
for $n=1,2,\cdots$.
Thus,  for any variable-length encoder, we have
\begin{align*}
\varepsilon_n\left( \varphi_n,na + \sqrt{n}L_1 \right)& \geq \Pr\left\{ P_{X^n}(X^n) \leq z_n K^{ - \alpha_c(na - \sqrt{n}L_1)} \right\} - z_n \\
& =  \Pr\left\{ \frac{1}{\sqrt{n}\alpha_c} \log \frac{1}{P_{X^n}(X^n)} \geq  \sqrt{n}a + L_1 - \frac{\log z_n}{\sqrt{n}\alpha_c} \right\} - z_n \\
& = \Pr\left\{ \frac{-\log P_{X^n}(X^n) - n\alpha_ca }{\sqrt{n}\alpha_c} \geq L_1 - \frac{\log z_n}{\sqrt{n}\alpha_c}\right\} - z_n 
\end{align*}
Set $z_n = K^{-\sqrt{n} \gamma}$, where $\gamma >0$ is a small constant that satisfies
\begin{equation} \label{eq:6-2-1}
L_1 + \frac{\gamma}{\alpha_c} < \inf \left\{ L \left| F_a(L) \leq \varepsilon \right. \right\}.
\end{equation}
Here, since we assume (\ref{eq:6-2-0}), it is obvious that there exists $\gamma >0$ satisfying the above inequality.
Then, we have
\begin{align*}
\varepsilon_n\left( \varphi_n,na + \sqrt{n}L_1 \right) \geq \Pr\left\{ \frac{-\log P_{X^n}(X^n) - n\alpha_c a }{\sqrt{n}\alpha_c} \geq L_1 + \frac{\gamma}{\alpha_c} \right\} - K ^{ -\sqrt{n}\gamma}.
\end{align*}
This implies that 
\begin{align*}
\limsup_{n \to \infty}\varepsilon_n\left( \varphi_n,na + \sqrt{n}L_1 \right) & \geq  \limsup_{n \to \infty}  \Pr\left\{ \frac{-\log P_{X^n}(X^n) - n\alpha_ca }{\sqrt{n}\alpha_c} \geq L_1 + \frac{\gamma}{\alpha_c} \right\} \nonumber \\
& >  \varepsilon,
\end{align*}
where the last inequality is derived from (\ref{eq:6-2-1}) and the definition of $F_a(L)$.

On the other hand, since we assume that $L_1$ is $(\varepsilon,a)$-achievable overflow threshold, it holds that
\begin{equation*}
\limsup_{n \to \infty} \varepsilon_n\left( \varphi_n,na + \sqrt{n}L_1 \right) \leq \varepsilon.
\end{equation*}
This is a contradiction.
Therefore, the proof of converse part has been completed. 
\end{IEEEproof}
%
%
%
%
%
%
%
%
%
\subsection{Computation for i.i.d. Sources}
Theorem \ref{theorem2} is a quite general result, because there is no restriction about the probability structure for the source.
However, to compute the function $L\left( \varepsilon,a|{\bf X}\right)$ is hard in general.
Next, we consider a simple case such as an i.i.d. source with countably {\it infinite} alphabet and we address the above quantity explicitly.

For an i.i.d. source, from Remark \ref{remark:5-2}, we are interested in $L\left( \left. \varepsilon,\frac{1}{\alpha_c} H(X)\right|{\bf X}\right) $.
To specify this quantity  for an i.i.d. source, we need to introduce the variance of self-information as follows:\begin{equation*}
\sigma^2 \stackrel{\mathrm{def}}{=} E\left( \left(-\log P_{X}(X) - H(X)\right)^2 \right),
\end{equation*}
where $H(X)$ is the entropy of the i.i.d. source defined by
\[H(X) = \sum_{{ x} \in {\cal X}} P_{X}({x}) \log \frac{1}{P_{X}({ x})}.\]
Here, we assume that the above variance exists.
Then, from Theorem \ref{theorem2} we obtain the following theorem.
\begin{theorem} \label{theorem3}
For any i.i.d. source, it holds that
\begin{eqnarray*} 
L\left( \left. \varepsilon,\frac{1}{\alpha_c} H(X) \right|{\bf X} \right) = \frac{1}{{\alpha_c}}\sigma \Phi^{-1}(1 - \varepsilon),
\end{eqnarray*}
where $\Phi^{-1}$ denotes a inverse function of $\Phi$ and 
$\Phi(T)$ is the Gaussian cumulative distribution function with mean $0$ and variance $1$, that is, $\Phi(T)$ is given by
\begin{eqnarray} \label{eq:5-1}
\Phi(T) & = & \int_{-\infty}^{T} \frac{1}{\sqrt{2\pi }}\exp\left[ -\frac{y^2}{2} \right] dy.
\end{eqnarray}
\end{theorem}
\begin{IEEEproof}
From the definition of $F_a(L)$,  we have
\begin{align*}
F_{H(X)/\alpha_c}(L) & =  \limsup_{n \to \infty} \Pr \left\{ \frac{ - \log P_{X^n}(X^n) -n H(X) }{\sqrt{n}\alpha_c } \geq  {L} \right\} \\
& =  \limsup_{n \to \infty} \Pr \left\{ \frac{ - \log P_{X^n}(X^n) - nH(X) }{\sqrt{n}\sigma}\geq  \frac{L \alpha_c }{\sigma} \right\} .
\end{align*}
On the other hand, since we consider the i.i.d. source
, from the asymptotic normality (due to the central limit theorem) it holds that
\begin{equation*}
 \lim_{n \to \infty} \Pr\left\{ \frac{\!-\! \log P_{X^n}(X^n)\!-\!nH(X) }{\sqrt{n }\sigma}\leq U \right\} =  \int_{-\infty}^{U} \frac{1}{\sqrt{2\pi }}\exp\left[ -\frac{y^2}{2} \right] dy.
\end{equation*}
This means that
\begin{align*}
F_{H(X)/\alpha_c}(L) = \int^{+\infty}_{\frac{L \alpha_c}{\sigma}} \frac{1}{\sqrt{2\pi }}\exp\left[ -\frac{y^2}{2} \right] dy.
\end{align*}
Thus, $L\left( \left. \varepsilon,\frac{1}{\alpha_c}H(X) \right| {\bf X} \right) $ is given by
\begin{align*} 
L\left( \left. \varepsilon,\frac{1}{\alpha_c}H(X) \right|{\bf X} \right)
& =  \inf \left\{ L \left| \int^{+\infty}_{\frac{L\alpha_c}{\sigma}} \frac{1}{\sqrt{2\pi }}\exp\left[ -\frac{y^2}{2} \right] dy  \leq \varepsilon \right. \right\} \\
& = \inf \left\{ L \left| 1- \Phi \left( \frac{L\alpha_c}{\sigma} \right) \leq \varepsilon \right. \right\}
\end{align*}
Since $\Phi \left( \frac{L\alpha_c}{\sigma} \right)$ is a continuous function and monotonically increases as $L$ increases, we have
\[
\frac{L\left( \left. \varepsilon,\frac{1}{\alpha_c}H(X) \right|{\bf X} \right) \alpha_c}{\sigma} = \Phi^{-1}(1-\varepsilon).
\]
Therefore, the proof has been completed. 
\end{IEEEproof}
\begin{remark}
As shown in the proof, the derivation of Theorem \ref{theorem3} is based on of the asymptotic normality of self-information. This means that the similar argument is valid for any source for which the asymptotic normality of self-information holds such as Markov sources (see, Hayashi \cite{Hayashi}).
\end{remark}
%
%
\subsection{Computation for Mixed Sources}
In this subsection we consider mixed sources.
The class of mixed sources is very important, because all of stationary sources can be regarded as forming mixed sources obtained by mixing stationary ergodic sources with respect to appropriate probability measures. Notice here that, in general, the mixed source does not have the asymptotic normality of self-information. So, we can not simply apply Theorem \ref{theorem3}.

The second-order achievable rates for mixed sources has been first considered by Nomura and Han \cite{NH2011} in the fixed-length source coding problem. In this paper, we also use the similar approach. The result in this subsection is analogous to the result in \cite{NH2011}.

We consider a mixed source consists of two stationary memoryless sources ${\bf X}_{i}
 = \{ X^n_{i} \}_{n=1}^\infty$ with $i=1,2$.
Then, the mixed source ${\bf X} = \{ X^n \}_{n=1}^\infty$ is defined by
\begin{equation} \label{mixed}
P_{X^n}({\bf x}) = w(1)P_{X_{1}^n}({\bf x}) + w(2)P_{X_{2}^n}({\bf x}),
\end{equation}
where $w(i)$ are  constants satisfying $w(1) + w(2) = 1$ and $w(i) > 0$ $(i=1,2)$.
Since two i.i.d. sources ${\bf X}_i$ $(i =1,2)$ are completely specified by giving just the first component $X_i$ $(i =1,2)$, we may write simply as ${\bf X}_i = \{X_i \}$ $( i=1,2)$ and define the variances:
\begin{eqnarray*}
\sigma^2_i = E \left( \log \frac{1}{P_{X_i}(X_i)}- H(X_i) \right)^2 \ (i=1,2) ,
\end{eqnarray*}
where we assume that these variances are exist, and define the entropy by
\[H(X_i) = \sum_{{ x} \in {\cal X}} P_{X_i}({x}) \log \frac{1}{P_{X_i}({ x})}.\]

Before showing second-order analysis we shall consider the first-order case.
Without loss of generality, we assume that $H(X_1) \geq H(X_2)$ holds.
\begin{theorem} \label{theo:4-1} For any mixed source defined by (\ref{mixed}), we have
\begin{align} R(\varepsilon|{\bf X}) = \left\{\begin{array}{ll}
\frac{H(X_1)}{\alpha_c} & \mbox{ if } 0 \leq \varepsilon < w(1),  \\
\frac{H(X_2)}{\alpha_c} & \mbox{ if } w(1) \leq \varepsilon < 1.
\end{array} \right.
\end{align}
\end{theorem}
\begin{IEEEproof}
This theorem can be obtained as an immediate consequence of Theorem \ref{theorem1} 
(or Theorem \ref{theo:0-1} combined with \cite[Example 1.6.1]{Han}).
\end{IEEEproof}

By using the above theorem we shall compute the second-order case.
As we have mentioned in the above, the asymptotic normality of self information does not hold for mixed sources. However, since we consider the case where ${\bf X}_i = \{X_i \}$ $( i=1,2)$ is an i.i.d. source, the following asymptotic normality  holds for each component i.i.d. source:
\begin{equation} \label{normality2}
\lim_{n \rightarrow \infty}\Pr\left\{ \frac{-\log P_{X_i^n}(X_i^n) - nH(X_i)}{\sqrt{n }{\sigma_i}} \leq U \right\}
 =  \Phi(U).
\end{equation}

The following lemma plays the key role in dealing with {\it mixed} sources in the proof of Theorem \ref{theorem4}.
\begin{lemma}[Han \cite{Han}, (see also {\cite[Lemma 4.1]{NH2011}})]\label{lemma}
Let $\{ z_n \}_{n=1}^\infty$ be any real-valued sequence.
Then for the mixed source ${\bf X}$ it holds that, for $i = 1,2$,
\[
\Pr\left\{ \frac{-\log P_{X^n}(X^n_{i}) }{\sqrt{n}}\geq z_n \right\}  \geq  \Pr\left\{ \frac{-\log P_{X^n_{i}}(X^n_{i}) }{\sqrt{n}} \geq z_n + \gamma_n \right\} - e^{-\sqrt{n}\gamma_n},
\]
\[
\Pr\left\{ \frac{-\log P_{X^n}(X^n_{i}) }{\sqrt{n}}\geq z_n \right\} \leq \Pr\left\{ \frac{-\log P_{X^n_{i}}(X^n_{i}) }{\sqrt{n}} \geq z_n - \gamma_n \right\} , 
\]
where $\gamma_n > 0$ satisfies $\gamma_1 > \gamma_2 > \cdots > 0,$ $\gamma_n \to 0$, $\sqrt{n} \gamma_n \to \infty$.
\end{lemma}

In the sequel, we consider the case that $0 \leq \varepsilon < 1$ and $w(1) \neq \varepsilon$ hold, because if $w(1) \neq \varepsilon$ holds the second-order achievable overflow threshold is trivial (cf. \cite[Remark 5.2]{NH2011}).
Then, given $ \varepsilon$ we classify the problem into three cases. We also assume that $H(X_1) \geq H(X_2)$ holds without loss of generality:

\begin{description}
\item[I] $H(X_1) = H(X_2)$ holds.
\item[II] $H(X_1) > H(X_2)$ and $w(1) > \varepsilon$ hold.
\item[III]	$H(X_1) > H(X_2)$ and $w(1) < \varepsilon$ hold.
\end{description}
In Case I, we shall compute $L\left(\left.\varepsilon,\frac{1}{\alpha_c}H(X_1)\right|{\bf X}\right)$ (this is equal to $L\left(\left.\varepsilon,\frac{1}{\alpha_c}H(X_2)\right|{\bf X}\right)$). In Case II and Case III we shall show $L\left(\left.\varepsilon,\frac{1}{\alpha_c}H(X_1)\right|{\bf X}\right)$ and  $L\left(\left.\varepsilon,\frac{1}{\alpha_c}H(X_2)\right|{\bf X}\right)$, respectively. 
Then, from Theorem \ref{theorem2} we obtain the following theorem:
\begin{theorem} \label{theorem4}
For any mixed source, it holds that
\begin{description}
\item[Case I]
\begin{eqnarray*} 
L\left( \left. \varepsilon,\frac{1}{\alpha_c} H(X_1) \right|{\bf X} \right) = T_1,
\end{eqnarray*}
where $T_1$ is specified by
\begin{eqnarray*} 
\varepsilon =  1 - \sum_{i=1}^2 w(i) \Phi \left( \frac{T_1\alpha_c}{\sigma_i} \right).
\end{eqnarray*}
\item[Case II]
\begin{eqnarray*} 
L\left( \left. \varepsilon,\frac{1}{\alpha_c} H(X_1) \right|{\bf X} \right) = T_2,
\end{eqnarray*}
where $T_2$ is specified by
\begin{eqnarray*} 
\varepsilon=  w(1) \left( 1- \Phi\left( \frac{\alpha_c T_2 }{\sigma_1}\right)\right).
\end{eqnarray*}

\item[Case III]
\begin{eqnarray*} 
L\left( \left. \varepsilon,\frac{1}{\alpha_c} H(X_2) \right|{\bf X} \right) = T_3,
\end{eqnarray*}
where $T_3$ is specified by
\begin{eqnarray*} 
\varepsilon =  w(1) + w(2)\left( 1 - \Phi\left( \frac{\alpha_c T_3}{\sigma_2}\right)\right).
\end{eqnarray*}
\end{description}
\end{theorem}
\begin{IEEEproof}
This theorem can be shown substantially same with \cite[Theorem 5.1]{NH2011}. We only show the proof of Case I in Appendix.
\end{IEEEproof}
\begin{remark}
In \cite{NH2011}, the countably infinite mixture of i.i.d. sources and general mixture of i.i.d. sources are  treated.
We can also obtain the infimum of $(\varepsilon,a)$-achievable overflow threshold in these cases by using the similar argument.
\end{remark}
%
%
%
%
%
%
%
%
%
%
\section{Concluding Remarks}
We have so far dealt with the overflow probability of variable-length coding with codeword cost. 
The overflow probability is important not only from the theoretical viewpoint but also from the engineering point of view.
As shown in the proofs of the present paper, the {\it information-spectrum} approach is substantial in the analysis of the overflow probability of variable-length coding. 

In particular, Lemma \ref{lemma1} and Lemma \ref{lemma2} are key lemmas.
The infimum of {\it first-order} achievable threshold and the infimum of {\it second-order} achievable threshold have been derived from these lemmas. 
Theorem \ref{theo:0-1} is also useful, because it enables us to apply results derived in the fixed-length coding problem, into the variable-length coding problem.

Finally, we shall note a generalization of the cost function. Although we only consider the memoryless cost function, all the results in this paper are valid for wide class of cost function as follows.
Let us define the cost function $c: {\cal U}^{\ast} \to (0, +\infty)$ considered in this paper.
The cost $c(u^l)$ of a sequence $u^l \in {\cal U}^l$ is defined by
\[
c(u^l) = \sum_{i=1}^{l}c(u_i|u_1^{i-1}),
\]
where $c(u_i|u_1^{i-1})$ is a conditional cost of $u_i$ given $u_1^{i-1}$ such that 
$0 < c(u_i|u_1^{i-1}) < \infty$ ($\forall i,$$\forall u_i \in {\cal U},$$\forall u_1^{i-1} \in {\cal U}^{i-1}$). 
The conditional {\it cost capacity} $\alpha_c(u_1^{i-1})$ given $u_1^{i-1}$ is defined by the positive unique root $\alpha$ of the equation
\begin{equation*}
\sum_{u_i \in {\cal U}} K^{-\alpha c(u_i|u_{1}^{i-1})} =1.
\end{equation*}
Furthermore, we assume that the conditional {\it cost capacity} $\alpha_c(u_1^{i-1})$ is independent on $u_1^{i-1}$, more exactly, $\alpha_c(u_1^{i-1}) = \alpha $ holds for all $u_1^{i-1} \in {\cal U}^{i-1}$.
Such a class of cost function has been first considered in \cite{HK97}. Han and Uchida \cite{Han00:cost} also have treated this type of cost function. 
Since the conditional cost capacity $\alpha_c(u_1^{i-1})$ is independent on $u_1^{i-1}$, 
all the results in this paper can be proved for this type of cost function. 
\appendices
\section{Proof of Theorem \ref{theo:scp}}
\renewcommand{\theequation}{A.\arabic{equation}}
\setcounter{equation}{0}
The proof consists of two parts. \\
(Sufficiency) We assume that (\ref{eq:scp}) holds. 
Set $R = R(0|{\bf X}) - \frac{2\gamma}{\alpha_c}$, where $\gamma >0$ is an arbitrary constant.
Then, from Corollary \ref{coro1} it holds that 
\[
R = \frac{1}{\alpha_c} \overline{H}({\bf X}) - \frac{2 \gamma}{\alpha_c}
\]
On the other hand, from Lemma \ref{lemma2} with $\eta_n = nR$, it holds that
\begin{align*}
\varepsilon_n(\varphi_n, nR)
& >  \Pr\left\{ \frac{1}{n\alpha_c} \log \frac{1}{P_{X^n}(X^n)} \geq R -\frac{1}{n\alpha_c}\log z_n \right\} - z_n,
\end{align*}
for any sequence $\{z_n \}_{n=1}^\infty\  (z_i > 0, i=1,2,\cdots)$.
Let $z_n$ be as $K^{-{n}\gamma}$, then we have
\begin{align*}
\varepsilon_n(\varphi_n, nR)
& > \Pr\left\{ \frac{1}{n\alpha_c} \log \frac{1}{P_{X^n}(X^n)} \geq R + \frac{\gamma}{\alpha_c}  \right\} - K^{-{n} \gamma} \\
& = \Pr\left\{ \frac{1}{n\alpha_c} \log \frac{1}{P_{X^n}(X^n)} \geq  \frac{1}{\alpha_c}\overline{H}({\bf X}) - \frac{\gamma}{\alpha_c}  \right\} - K^{-{n} \gamma} \\
& = \Pr\left\{ \frac{1}{n\alpha_c} \log \frac{1}{P_{X^n}(X^n)} \geq  \frac{1}{\alpha_c}\underline{H}({\bf X}) - \frac{\gamma}{\alpha_c}  \right\} - K^{-{n} \gamma}.
\end{align*}
Noting that $\gamma > 0$ is a constant,  from the definition of $\underline{H}({\bf X})$, we have
\begin{eqnarray*}
\lim_{n \to \infty} \varepsilon_n(\varphi_n, nR) = 1.
\end{eqnarray*}
Therefore, the sufficiency has been proved. \\
%
%
(Necessity)
Set $R =  R(0|{\bf X}) - \gamma$ and $z_n = K^{-n\gamma}$, where $\gamma > 0$ is an arbitrary constant. 
From Lemma \ref{lemma1} with $\eta_n = nR$, there exists a variable-length encoder $\varphi_n$ that satisfies
\begin{eqnarray*}
\varepsilon_n(\varphi_n, nR) < \Pr\left\{ \frac{1}{n\alpha_c} \log \frac{1}{P_{X^n}(X^n)} \geq R - \frac{\gamma}{\alpha_c} \right\} +  K^{\alpha_c c_{max}+1}K^{-n\gamma},
\end{eqnarray*}
for $n=1,2,\cdots$.
Here, let $\varphi_n^{\ast}$ denote the variable-length coder satisfying the above.
Assuming that the source satisfies the strong converse property, we have
\begin{align*}
1 &=\ \liminf_{n \to \infty} \varepsilon_n(\varphi_n^{\ast}, nR)  \leq  \liminf_{n \to \infty} \Pr\left\{ \frac{1}{n\alpha_c} \log \frac{1}{P_{X^n}(X^n)} \geq R - \frac{\gamma}{\alpha_c} \right\} +\limsup_{n \to \infty}K^{-n\gamma + \alpha_c c_{max}+1} \\
& = \liminf_{n \to \infty} \Pr\left\{ \frac{1}{n\alpha_c} \log \frac{1}{P_{X^n}(X^n)} \geq R - \frac{\gamma}{\alpha_c} \right\}.
\end{align*}
This means that 
\begin{eqnarray*}
\lim_{n \to \infty} \Pr\left\{ \frac{1}{n\alpha_c} \log \frac{1}{P_{X^n}(X^n)} < R\right\} = 0.
\end{eqnarray*}
Thus, $R \leq \frac{1}{\alpha_c}\underline{H}({\bf X})$ holds from the definition of $\underline{H}({\bf X})$. On the other hand, from Corollary \ref{coro1}, it holds that
\begin{eqnarray*}
R = R(0|{\bf X}) - \gamma =  \frac{1}{\alpha_c}\overline{H}({\bf X}) - \gamma.
\end{eqnarray*}
Thus, we have $ \frac{1}{\alpha_c} \overline{H}({\bf X}) - \gamma \leq  \frac{1}{\alpha_c} \underline{H}({\bf X})$. Notice that $\gamma >0$ is arbitrarily, we have
\[
\frac{1}{\alpha_c}\overline{H}({\bf X}) \leq \frac{1}{\alpha_c}\underline{H}({\bf X})
\]
Hence, we have
$\overline{H}({\bf X}) = \underline{H}({\bf X})$.
Therefore, necessity has been proved.
\IEEEQED
\section{Proof of Theorem \ref{theorem4}}
\renewcommand{\theequation}{B.\arabic{equation}}
\setcounter{equation}{0}
We only show Case I. The proofs of Case II and Case III are similarly to that of Case I and \cite[Theorem 5.1]{NH2011}

From the definition of $F_a(L)$,  we have
\begin{align*}
F_{H(X_1)/\alpha_c}(L) & =  \limsup_{n \to \infty} \Pr \left\{ \frac{ - \log P_{X^n}(X^n) -n H(X_1) }{\sqrt{n}\alpha_c } \geq  {L} \right\} \\
& = \limsup_{n \to \infty}\sum_{i=1}^2\Pr \left\{ \frac{-\log P_{X^n}(X_i^n) - nH(X_1)}{\sqrt{n}\alpha_c} \geq L  \right\} w(i)
\end{align*}
The last equality is derived from the definition of the mixed source.
Thus, from Lemma \ref{lemma} we have
\begin{IEEEeqnarray}{rCl} \label{5-1}
\lefteqn{\sum_{i=1}^2 \limsup_{n \to \infty} \Pr \left\{ \frac{-\log P_{X_i^n}(X_i^n) - nH(X_1)}{\sqrt{n}\alpha_c} \geq L  -\gamma_n \right\} w(i)} \nonumber \\
& \geq & F_{H(X_1)/\alpha_c}(L) \nonumber \\ & \geq & 
 \sum_{i=1}^2 \liminf_{n \to \infty}\Pr \left\{ \frac{\!-\!\log P_{X_i^n}(X_i^n)\!-\!nH(X_1)}{\sqrt{n}\alpha_c}  \geq L +\gamma_n \right\}   w(i)
\end{IEEEeqnarray}
where $\gamma_n$ is specified in Lemma \ref{lemma}.

Then, noting that $H(Y_1)=H(Y_2)$ holds,  from the asymptotic normality, we have
\begin{align*}
\lim_{n \rightarrow \infty}\Pr\left\{  \frac{\!-\log P_{X_i^n}(X_i^n)- nH(X_1) }{\sqrt{n} \sigma_i} \geq \frac{L\alpha_c}{\sigma_i} \right\} & = \int^{\infty}_{\frac{L\alpha_c}{\sigma_i}} \frac{1}{\sqrt{2\pi}}\exp\left[ -\frac{z^2}{2} \right] dz \\
& = 1- \Phi \left( \frac{L\alpha_c}{\sigma_i} \right)
\end{align*} 
for $i=1,2$.
Noting  that $\gamma_n \to 0$ as $n \to \infty$ and the continuity of normal distribution function, we have
\begin{align*}
\sum_{i=1}^2 \limsup_{n \to \infty} \Pr \left\{ \frac{-\log P_{X_i^n}(X_i^n) - nH(X_1)}{\sqrt{n}\alpha_c} \geq L  -\gamma_n \right\} w(i) = 
1 - \sum_{i=1}^2 w(i) \Phi \left( \frac{L\alpha_c}{\sigma_i} \right).
\end{align*}
Similarly, the last term in (\ref{5-1}) is evaluated as 
\begin{align*}
\sum_{i=1}^2 \liminf_{n \to \infty} \Pr \left\{ \frac{-\log P_{X_i^n}(X_i^n) - nH(X_1)}{\sqrt{n}\alpha_c} \geq L  +\gamma_n \right\} w(i) = 
1 - \sum_{i=1}^2 w(i) \Phi \left( \frac{L\alpha_c}{\sigma_i} \right).
\end{align*}
Hence, we have proved Case I of the theorem.
\IEEEQED




\ifCLASSOPTIONcaptionsoff
  \newpage
\fi



\bibliographystyle{IEEEtran}

%

%








\end{document}